\documentclass[prx,twocolumn,aps,epsf,showpacs,superscriptaddress,longbibliography]{revtex4-1}
\usepackage[pdftex]{graphicx}

\usepackage{dcolumn}
\usepackage{bm}
\usepackage{epsfig}
\usepackage{latexsym} 
\usepackage{amsmath}
\usepackage{amssymb}
\usepackage[usenames, dvipsnames]{color}
\usepackage{array}
\usepackage{bbm}
\usepackage{hyperref}
\hypersetup{
	pdftoolbar=true,        		 
	pdfmenubar=true,     		 
	pdffitwindow=false,     	 
	pdfstartview={FitH},    	 
	pdftitle={},   
	pdfauthor={},     		 
	pdfsubject={},   			 
	pdfcreator={},   	        	   	 
	pdfproducer={}, 		
	pdfnewwindow=true,      	
	colorlinks=true,       		
	linkcolor=black,          	
	citecolor=blue,        		
	filecolor=magenta,    		
	urlcolor=blue          		
}

\usepackage{color}
\usepackage{array}
\usepackage{cancel}
\usepackage{ulem}

\usepackage{booktabs}
\newcolumntype{C}[1]{>{\centering\arraybackslash}m{#1}}
\AtBeginDocument{
\heavyrulewidth=.08em
\lightrulewidth=.05em
\cmidrulewidth=.03em
\belowrulesep=.65ex
\belowbottomsep=0pt
\aboverulesep=.4ex
\abovetopsep=0pt
\cmidrulesep=\doublerulesep
\cmidrulekern=.5em
\defaultaddspace=.5em
}
\usepackage{booktabs}
\newcolumntype{R}[1]{>{\raggedleft\arraybackslash}p{#1}}
\AtBeginDocument{
\heavyrulewidth=.08em
\lightrulewidth=.05em
\cmidrulewidth=.03em
\belowrulesep=.65ex
\belowbottomsep=0pt
\aboverulesep=.4ex
\abovetopsep=0pt
\cmidrulesep=\doublerulesep
\cmidrulekern=.5em
\defaultaddspace=.5em
}

\newcommand{\<}{\langle}

\renewcommand{\>}{\rangle}
\renewcommand{\(}{\left(}
\renewcommand{\)}{\right)}
\renewcommand{\[}{\left[}
\renewcommand{\]}{\right]}

\renewcommand{\d}{\partial}

\newcommand{\gs}[1]{\textcolor{Green}{\sout{#1}}}

\renewcommand{\gs}[1]{{}}

\newcommand{\subheader}[1]{\vspace{4pt}\noindent{\bf #1 --}}

\begin{document}
\title{Tracking the quantized information transfer at the edge of a chiral Floquet phase}
\author{Blake R. Duschatko}
\affiliation{Department of Physics, University of Texas at Austin, Austin, TX 78712, USA}
\author{Philipp T. Dumitrescu}
\affiliation{Department of Physics, University of Texas at Austin, Austin, TX 78712, USA}
\author{Andrew C. Potter}
\affiliation{Department of Physics, University of Texas at Austin, Austin, TX 78712, USA}

\begin{abstract}
Two-dimensional arrays of periodically driven qubits can host inherently dynamical topological phases with anomalous chiral edge dynamics. These chiral Floquet phases are formally characterized by a dynamical topological invariant, the chiral unitary index. Introducing a quantity called the chiral mutual information, we show that this invariant can be precisely interpreted in terms of a quantized chiral transfer of quantum information along the edge of the system, and devise a physical setup to measure it.
\end{abstract}

\maketitle
Time-periodic (Floquet) driving enables a
host of symmetry-breaking and topological phases with inherently dynamical properties that could not occur in static or equilibrium settings~\cite{kitagawa2010topological,rudner2013anomalous,po2016chiral,harper2016stability,po2017radical,titum2016anomalous,jiang2011majorana,von2016phaseI,else2016classification,potter2016topological,roy2016abelian,roy2016periodic}. A striking set of examples are Chiral Floquet (CF) phases of driven $2d$ systems, which exhibit trivial bulk dynamics, but whose edges act as unidirectional ``conveyer belts" for quantum states~\cite{kitagawa2010topological,rudner2013anomalous,titum2016anomalous,po2016chiral,harper2016stability,po2017radical}. CF phases were first theoretically constructed in a non-interacting fermion model~\cite{kitagawa2010topological,rudner2013anomalous,titum2016anomalous}, and were subsequently generalized to interacting bosonic~\cite{po2016chiral,harper2016stability}, fermionic~\cite{po2016chiral,fidkowski2017interacting}, and fractionalized (anyonic)~\cite{po2017radical} Floquet systems.

Despite superficial similarities to the more familiar quantum Hall effect, CF phases represent a distinct and intrinsically dynamical topological phenomena. Namely, CF phases have vanishing Chern number, and are instead governed by a dynamical topological invariant -- the chiral unitary index~\cite{po2016chiral}, $\nu$. For non-fractionalized phases (i.e. without topological order and anyonic excitations), the index takes the form of the logarithm of a positive rational fraction, $\nu \in \log \mathbb{Q}_+$~\cite{gross2012index,po2016chiral}, and can be heuristically interpreted as the (log of the) ratio of the number of quantum states transferred to the right divided by the number transferred to the left by the edge dynamics, during each period.  Formally, this index has been defined in terms of abstract observable algebras~\cite{po2016chiral,gross2012index} and matrix product operator methods~\cite{po2016chiral, burak2017mpo-lpu, cirac2017mpo-topo-inv}, enabling a rigorous classification of $2d$ Floquet topology in the absence of symmetry and intrinsic topological order.

Notwithstanding its theoretical utility as a formal tool for characterizing $2d$ Floquet topological phases, this algebraic formulation is both physically opaque and not amenable to experimental measurement. In this paper, we address both of these shortcomings by showing that it is possible to reformulate $\nu$ in terms of a chiral imbalance in the transfer of quantum information. Our construction not only yields a physically transparent interpretation of the chiral unitary index, but also enables a realistic scheme to measure this dynamical topological invariant using existing experimental techniques. 

Our strategy will be to introduce additional non-dynamical ancillary qubits that are initially locally entangled with the edge, and serve as ``tracers" to track the dynamics of entanglement within the system during its evolution.  We show that this setup enables us to recast the chiral unitary index in terms of an appropriate chiral combination of mutual information between ancilla and system subregions. This chiral mutual information, ($\chi$MI), can be constructed from any extensive entanglement measure, including Renyi entropies, which can be measured experimentally via ``SWAP"-based many-body quantum interference~\cite{horodecki2002method,islam2015measuring}.

In contrast to recently proposed measurement schemes based on observing quantized magnetization~\cite{nathan2017quantized} in charge-conserving systems, our setup does not require any extraneous symmetries or conservation laws, which are both inessential to the underlying topological dynamics and also typically absent in the qubit systems that are most likely to realize these CF dynamics.  Moreover, our proposal avoids the use of external leads, which, while natural for electronic materials, are difficult to synthesize for atomic or qubit systems.

\subheader{Setup}
In interacting settings, care is required to avoid drive-induced heating which would lead to a highly entangled incoherent state and destroy any Floquet topology. To this end, two strategies have emerged. First, heating can be postponed to exponentially long times by rapid driving~\cite{abanin2017effective,kuwahara2016floquet,else2017prethermal}. Alternatively, heating can be 
prevented~\footnote{While MBL has been firmly established in $1d$ systems~\cite{imbrie2016many}, its stability to rare-region effects in higher dimensional systems has been questioned~\cite{de2017stability}. While an important point of principle these effects are practically irrelevant, as they occur on time-scales that are doubly exponentially long in the disorder strength, which can easily be made to exceed any practical experimental lifetime (or even the age of the universe!) for moderate disorder strength. Moreover, these rare-region worries can likely be side-stepped by implementing quasi-periodic ``disorder" potentials, which do not have arbitrarily rare regions~\cite{khemani2017two}.}  
by applying random disorder to produce many-body localization (MBL)~\cite{nandkishore2015many,altman2015universal,lazarides2015fate,ponte2015many}.

In what follows, we will consider only rational CF phases, whose bulk Floquet evolution is trivial (i.e. lacks topological order) and MBL. 
In this setting, it is always possible to decompose the time-evolution operator for one period, or Floquet operator, into bulk and edge components~\cite{po2016chiral}:
 \begin{align}
 U_F = \mathcal{T}\{e^{-i\int_0^TH(t)dt}\} =  U_\text{edge}e^{-iH_\text{bulk}T}.
 \label{eq:Uedge}
 \end{align}
Here $H_\text{bulk}$ is a trivial, static MBL Hamiltonian, and $U_\text{edge}$ is an effective $1d$ evolution that acts only on qubits within a few localization lengths from the boundary. The unitary $U_\text{edge}$ is anomalous, in the sense that it cannot be generated by any local edge Hamiltonian $H_{1d}(t)$, even though it acts locally on the edge qubits. A precise construction of Eq.~\ref{eq:Uedge} is reviewed in Appendix~\ref{app:formal}.

Although our constructions generalize to generic rational CF phases, for concreteness, we will  frame our discussion in terms of the simplest case of a $2d$ qubit array with $\nu=\log 2$. The topological edge dynamics in this case will be equivalent to a clockwise translation of the state of each edge qubit to the right neighboring site along the edge. Initially, we will  ignore the trivial MBL dynamics of the bulk qubits and formulate a method to extract the topological dynamics encoded in $U_\text{edge}$. Subsequently, we explain a method to experimentally extricate these topological entanglement signatures from those arising from trivial bulk MBL dynamics. 

\begin{figure}[t]
\centering
\includegraphics[width=0.8\columnwidth]{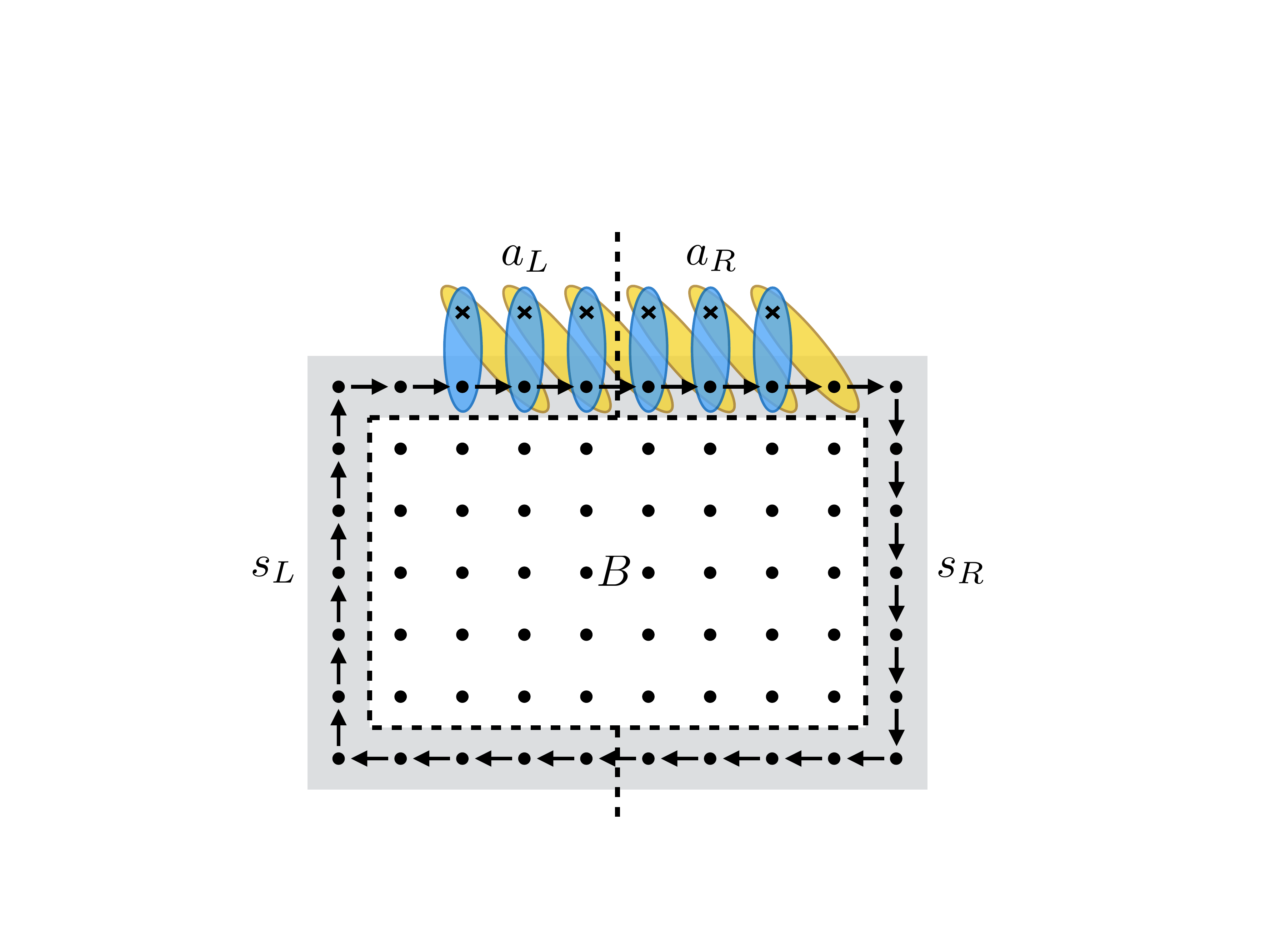}
\caption{
{\bf System Schematic --}
CF system made from qubits (dots) on a square lattice. Ancilla qubits (crosses) are initially entangled with corresponding edge qubits (blue bonds). During a Floquet period, the qubits in the bulk undergo a trivial MBL evolution (not shown), while qubits at the edge (gray shading) undergo a net translation -- shifting the system-ancilla entanglement (yellow bonds). Dashed lines indicate entanglement cuts that divide the system edge (s) and ancilla (a) into left (L) and right (R) regions, and separate edge and bulk (B). Generically the edge region (gray shaded) can contain multiple layers of spins, and should be thick enough to include several localization lengths.
}
\label{fig:schematic}
\end{figure}

 \subheader{Chiral mutual information} To relate the chiral unitary index of the edge dynamics described by $U_\text{edge}$ to a quantized chiral information transfer along the edge, we need a way to track the flow of quantum information. Here, we face a challenge: unlike charge or heat flow at a quantum Hall edge, quantum information is not associated with a conserved quantity carried by a locally measurable current. 

To circumvent this difficulty, we introduce ancillary qubits that do not participate in the system dynamics, but serve as initial location ``tags" for the information stored in the system spins. Specifically, we take an even number, $N_a$, of ancillary qubits lined up with edge qubits in the interval $[-N_a/2+1  ,N_a/2]$, each initially in a maximally entangled singlet state with its partner system qubit (Fig.~\ref{fig:schematic}). The remaining system qubits not paired with an ancilla are initialized into any convenient short-range entangled state.

Evolving the system by one period of the CF evolution to state: $|\psi(T)\>=U_\text{edge}|\psi_0\>$, produces a chiral transfer of states in the system. The resulting quantum information transfer from this CF dynamics can be extracted by dividing the system and ancilla spins into left (L) and right (R), and measuring the difference between mutual information between the left ancillas and right spins compared to that between the right ancillas and left spins:
\begin{align}
\chi = \mathcal{I}(a_L,s_R)-\mathcal{I}(a_R,s_L).
\label{eq:chi}
\end{align}
We will henceforth refer to this quantity as the chiral mutual information ($\chi$MI). Here, $\mathcal{I}(A,B) = \[S(A)+S(B)-S(A\cup B)\] /2$ is the mutual information between regions $A$ and $B$ in the state $|\psi(T)\>=U_\text{edge}|\psi_0\>$, and $S(A) = -\text{tr}\(\rho_A\log\rho_A\)$ is the entanglement entropy of the reduced density matrix in region $A$. 

For the ideal case of pure chiral translation, $U_\text{edge}=\tau$, precisely one singlet of entanglement crosses the left-right entanglement cut, so that $\chi = \nu = \log 2$. For a generic edge evolution, entanglement also spreads in a non-universal fashion in addition to this chiral shift. However, we will next show that in the limit of large system size and ancilla number that the $\chi$MI remains precisely quantized to $\chi = \nu$.

\begin{figure*}[t]
\includegraphics[width=\textwidth]{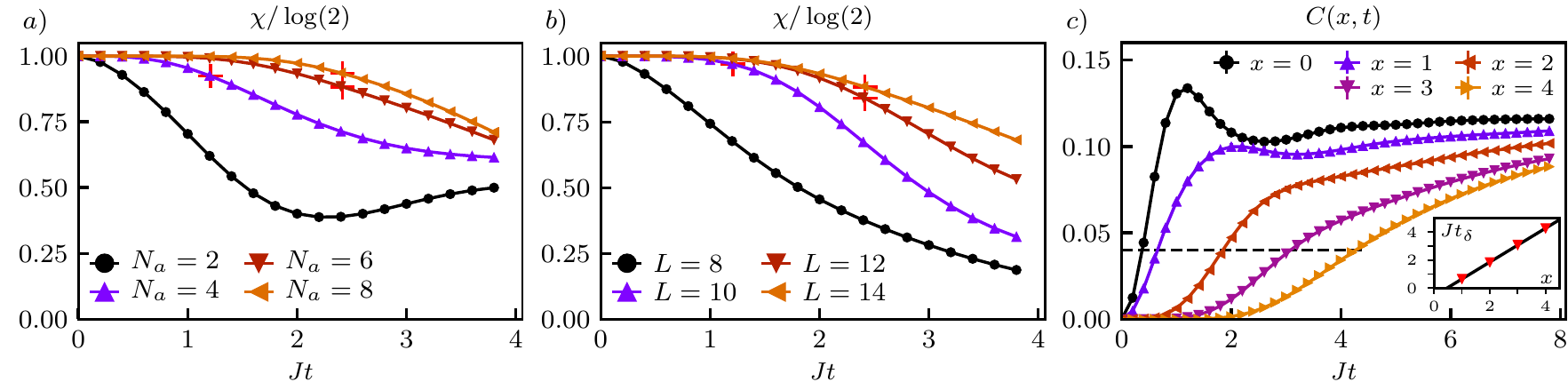}
\caption{{\bf Finite size dependence of chiral mutual information, $\chi(t)$ --}  normalized to the ideal value, $\log 2$, for a fixed random initial state and disorder realization, for (a) fixed ancilla number $N_a=6$, and various system sizes $L$, and (b) fixed $L$ and varying $N_a$. (c) Behavior of the butterfly correlator, $C(x,t)$ for $L=10$ spins averaged over all initial states, and $100$ disorder realizations. Inset shows a linear fit to the butterfly velocity as described in the main text. Note, that the similarity between $L=12,14$ in (a) and for $N=6,8$ in (b) arises since these geometries having the same minimal distance, $\ell_a$, due to periodic boundary conditions. 
}
\label{fig:numerics}
\end{figure*}

\subheader{Quantization of the chiral mutual information}
We now relate $\chi$ to the chiral unitary invariant $\nu$, beyond the special case discussed above. As a preliminary step, we note that for a generic CF evolution with $\nu = \log 2$, we can generically parameterize the edge evolution as
\begin{align}
U_\text{edge} = e^{-iH_{1d}t}\tau,
\label{eq:uedge}
\end{align}
where $\tau$ is an operator which translates each qubit's state one site to the right,
$H_{1d}$ is a local $1d$ Hamiltonian, and $t$ is a parameter with units of time (see Appendix~\ref{app:formal}). 
We further note that, since we are presently considering the effect of $U_\text{edge}$, which does not mix bulk and edge degrees of freedom, we can use the complementarity property of entanglement to simplify Eq.~\ref{eq:chi} to:
\begin{align}
\chi_\text{edge} = S(s_R)-S(s_L)
\label{eq:chiedge}
\end{align}

Our task is then to show that the non-topological edge evolution produced by $H_{1d}$  transfers equal amounts of quantum information to $s_L$ and $s_R$, leaving $\chi$  invariant regardless of the form of $H_{1d}$. The local nature of $H_{1d}$ allows the decomposition: $H_{1d} = H_L+H_R+V$, where $H_{L/R}$ contain terms acting only on the left or right half of the entanglement cut, and $V$ includes interactions crossing the cut. 

Moreover, unlike the anomalous chiral translation $\tau$, the evolution under $H_{1d}$ can be decomposed into the product of many infinitesimal time-steps. This allows us to focus on a single infinite time step evolving from $t_0\rightarrow t_0+\Delta t$, where $0\leq t_0\leq t$ is some intermediate time, and take the limit of $\Delta t\rightarrow 0$. Here, to $\mathcal{O}(\Delta t^2)$, we can factorize: $U(\Delta t)\approx U_LU_Re^{-iV\Delta t}+\mathcal{O}(\Delta t^2)$. Since $U_{L/R} = e^{-iH_{L/R}\Delta t}$ act only on the $L/R$ sides of the entanglement cut, they do not effect the entanglement of either region. Hence, we must only consider $U_V(\Delta t)=e^{-iV\Delta t}$, which only effects spins within a finite subset of the ancilla-entangled region. 

Intuitively, if 
the region covered by ancillas is very large, then the evolution up to time $t_0$ cannot propagate unentangled degrees of freedom outside the ancilla region to the vicinity of the entanglement cut effected by $V$. Formally, this is guaranteed to accuracy $e^{-N_a/v_\text{LR}t}$ by causal bounds on local Hamiltonian evolution, where $v_\text{LR}$ is the Lieb-Robinson velocity~\cite{lieb1972finite}. Due to this local maximal entanglement, the  the effect of $U_V$ acting on the edge is equivalent to $U_V^T$ acting only on the ancilla qubits. However, applying a unitary operation to the ancillas cannot change the entanglement of the system spins, implying ${\d S(s_{L/R})}/{\d t}=0$. Together with the simplification Eq.~\ref{eq:chiedge}, this implies ${\d\chi_\text{edge}}/{\d t}=0$, i.e. that the $\chi$MI is generically precisely quantized and equal to the chiral unitary index in the large system size and ancilla number limit. 

While the above steps make use of the assumption that the ancilla and systems were initially maximally entangled, numerical simulations (Appendix~\ref{sec:partialEnt}) indicate that $\chi$ remains quantized for only partial system-ancilla entanglement, so long as each system-ancilla pair initially has the same amount of entanglement. 

\subheader{Numerical validation and finite-size corrections}
Away from the limit of infinite system size $L$ and ancilla number $N_a$, we expect finite-size induced deviations of $\chi$ from its quantized value. Intuitively, these arise when the non-topological parts of the edge evolution allow information from outside the ancilla covered region, which is not tracked in $\chi$, to propagate across the entanglement cut. Such processes propagate at a maximum speed $v_\text{LR}$ and must cover a minimum distance
\begin{align}
\ell_a = \text{min}\[\frac{N_a}{2}-1,\frac{L-N_a}{2}-1\],
\end{align}
were the second argument accounts for periodic boundary conditions. Thus for $t\ll \ell_a/v_\text{LR}$, $\chi$ should remain asymptotically close to its quantized value $\nu$.

To verify these simple estimates, we perform numerical simulations for a qubit chain of length $L$, evolving for a single Floquet period using $U_\text{edge}$ of the form Eq.~\ref{eq:uedge} with 
\begin{align}
H_{1d} = \sum_{i=0}^{L-1} \(J\vec{S}_i\cdot\vec{S}_{i+1}+h^z_iS_i^z+h^x_iS_i^x\).
\label{eq:H1d}
\end{align}
%
%
Here $\vec{S}_i = \frac{1}{2} \vec{\sigma}_i$ and the random fields are drawn from the uniform distribution $h^{x/z}_i \in [-J,J]$. For these parameters, $H_{1d}$ is  non-integrable, has no symmetries, and is thermalizing (the latter feature is not essential since the chiral translation $\tau$ would inevitably prevent localization even at strong disorder~\cite{po2016chiral}). We begin from an initial state in which the ancilla-region qubits form entangled singlets with their corresponding ancillas, and the remainder point along the z-axis of the Bloch sphere. Then, each qubit is rotated by a random angle $\theta_i$, drawn independently from $[0,2\pi)$ in the $z-y$ plane of the Bloch sphere, to produce a generic initial state without special features.

Figure~\ref{fig:numerics} shows $\chi$ as a function of the dimensionless parameter $Jt$, for various $L$ and $N_a$. In each case, $\chi$ initially is near its ideal quantized value $\log 2$, before developing systematic deviations. We observe that the near quantization generally persists over longer and longer time intervals as $\ell_a$ is increased, qualitatively agreeing with the intuition outlined above.  Note that in some cases, increasing $N_a$ or $L$ does not always increase $\ell_a$.

To establish a quantitative relationship between the deviation time and $\ell_a$, we extract an estimate for the Lieb-Robinson velocity by measuring the so-called ``butterfly" correlator~\cite{shenker2014black}:
\begin{align}
C(x,t) = -4 \overline{\< [e^{iH_{1d}t}S^z_xe^{-iH_{1d}t},S_0^z(0)]^2\>},
 \label{eq:butterfly}
\end{align}
where, $\overline{\(\dots\)}$ indicates an average over disorder configurations. Heuristically, $C(x,t)$ measures how perturbing a spin at site $0$ effects the measurement of a spin at site $x$ a time $t$ later, and hence, $C(x,t)$ is essentially zero until time $t \approx x/v_\text{LR}$. Thus, we can estimate $v_\text{LR}$ by identifying the time $t_\delta(x)$ where $C(x,t)$ reaches an arbitrary threshold $\delta \ll 1$, and performing a linear fit to extract: $v_\text{LR}\approx x/t_\delta(x)$.  Plotting the times where $v_\text{LR}t = \ell_a$ (red crosses) in Fig.~\ref{fig:numerics}, we observe quantitative agreement with the interpretation of the finite size errors presented at the beginning of this section.

\subheader{Effect of bulk MBL dynamics}
Having discussed $\chi$MI for the edge alone, we now consider the dynamics of the entire system including the trivial bulk MBL motion. This introduces two important changes. First, the topological edge motion does not occur precisely on the outermost row of qubits, but rather spreads into the bulk with an exponentially decaying envelop with characteristic length $\xi$, the localization length. This can be addressed by covering a fattened edge strip of width $W\gg \xi$ with ancillas, which captures the edge motion to accuracy $e^{-W/\xi}$. 

Second, the MBL dynamics can cause information to leak between the ancilla covered region at the edge into the bulk. Edge-bulk  entanglement generated far from the entanglement cut does not effect the mutual information terms in $\chi$,  however local cyclic motion of qubits around the triple intersection of regions $L$, $R$, and $B$ can produce non-topological contributions to $\chi$, even for a large edge strip $W$.

To extract the topological edge contribution, one can evolve the system for $n\gg 1$ Floquet periods. In so doing, the topological edge contributions accumulate linearly $\chi_\text{edge}= n\nu$, whereas the trivial bulk contributions are bounded by $\Delta\chi_\text{bulk}(nT) \lesssim \log nT$ due to the slow glassy MBL dynamics~\cite{bardarson2012unbounded}. Hence measuring the asymptotic slope, $\lim_{n\rightarrow\infty}\chi(nT)/n$ enables one to extract the topological edge contribution. Note that this procedure requires that the length of each edge region be much larger than  $v_\text{LR}nT$ and the width $W$ to be much larger than $\xi \log(nT)$ to prevent the edge spins undergoing the CF dynamics and the bulk spins from entangling with those beyond the ancilla-covered region during the time evolution.

\subheader{Experimental proposal} 
 In addition to providing a complementary, physically intuitive formulation of the chiral unitary invariant, the above construction reveals an experimental protocol to measure $\chi$.  While the von Neumann entropies in $\chi$ are challenging to measure directly, we note that Renyi versions of the $\chi$MI, denoted $\chi_n$, can be equally well formulated using the Renyi entropy $S_n(A) = \text{tr}(\rho_A^n )/(1-n)$,  for any index $n$, in place of von Neumann entropies in the mutual information terms in Eq.~\ref{eq:chi}. The proof of quantization of $\chi_n$ follows through as above and with the same restrictions. The second Renyi entropy ($n=2$) is particularly significant, as this quantity can be directly measured by making two copies of the system, and performing an interferometric measurement of the SWAP operator that exchanges the copies~\cite{horodecki2002method,islam2015measuring}.  

\begin{figure}
\centering
\includegraphics[width=\columnwidth]{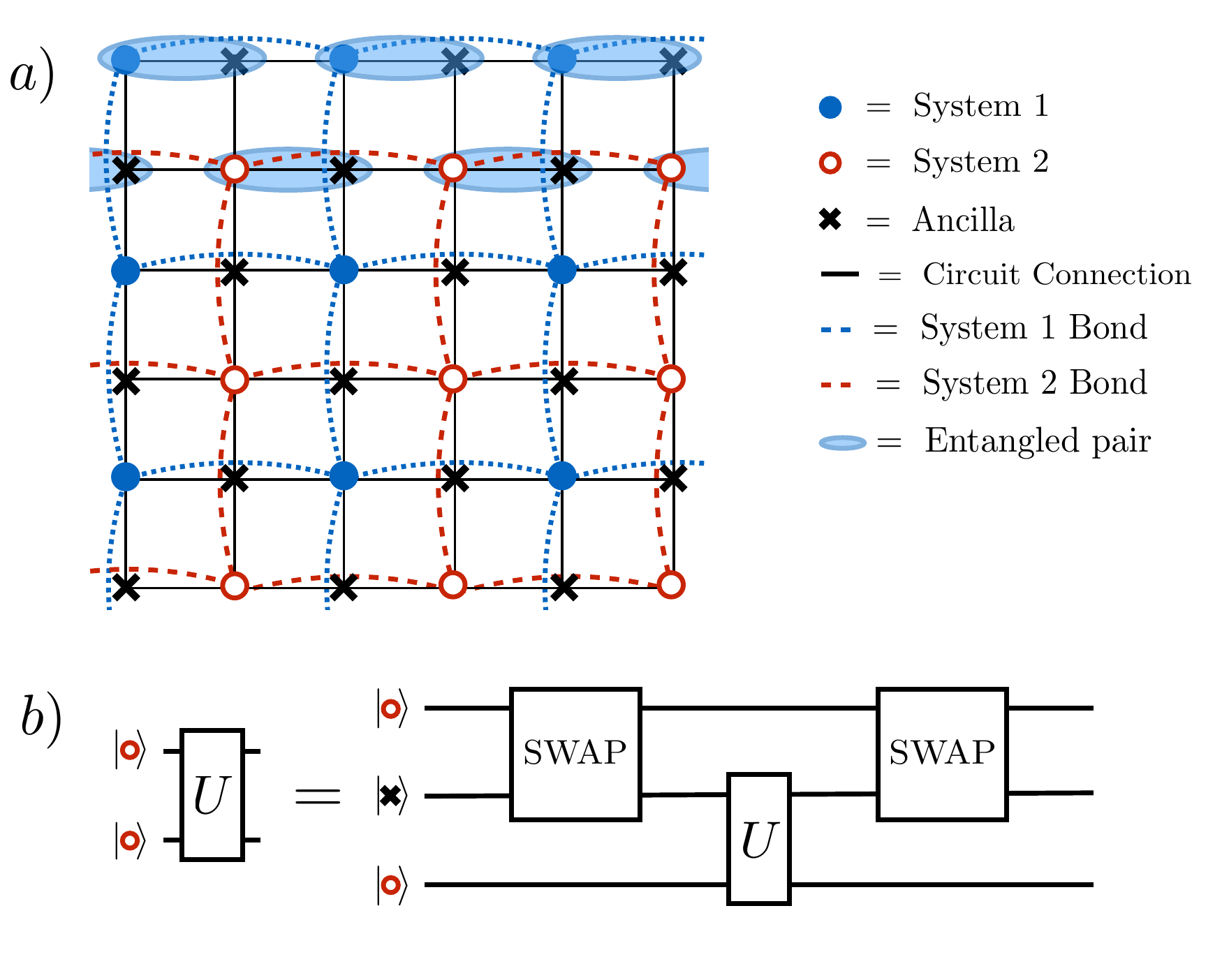}
\caption{
{\bf Experimental setup -- } (a) two copies of a chiral Floquet phase along with ancilla qubits can be implemented virtually in a $2d$ planar array of qubits with only nearest neighbor two-qubit gates. (b) A unitary gate, $U$, between spins in the same system can be implemented via two intermediate SWAP operations using an ancilla qubit.
}
\label{fig:experiment}
\end{figure}

This enables a direct measurement of each term in $\chi$, and opens the door to observing the quantized $\chi$MI in AMO systems such as $2d$ arrays of superconducting qubits, or trapped atomic, molecular, or ion systems. In particular, two dimensional qubit arrays have presently been fabricated by multiple groups. Furthermore, inductive couplings between the qubits are described by the effective Hamiltonian~\cite{roushan2017spectroscopic}:
\begin{align}
H_\text{2-qubit}=\sum_{\<i,j\>}\frac 12J_{i,j}(t)\(S^+_iS^-_j+S^+_iS^-_j\)
\end{align}
where $\<i,j\>$ denotes nearest neighbors. Applying this interaction for time $t = \frac{\pi}{4J_{i,j}}$ 
``hops" an excitation (logical $|1\>$) between nearest neighbor sites, and can be used to implement idealized SWAP models for the CF phase with $\nu=2^{\pm 1}$~\cite{po2016chiral}. Note that this XY-type interaction differs from Heisenberg type interactions, $\sim \vec{\sigma}_i\cdot\vec{\sigma}_j$, discussed in~\cite{po2016chiral} only by a $\sigma^z_i\sigma^z_j$ term that generates an unimportant conditional phase that is not required for the CF implementation.

To implement the above protocol to measure the $\chi$MI in these systems, we nominally need four distinct $2d$ layers: two copies each of the CF system and ancilla spins. While such a multilayer structure is impractical to directly fabricate, the layer structure can instead be ``virtually" implemented within a single physical qubit layer with nearest neighbor interactions as shown in Fig.~\ref{fig:experiment}a. In this setup, a four-site unit cell is used to implement the four distinct virtual layers. To synthesize a two-qubit unitary operation $U$ between virtual neighbors in one copy of the system (which are next-nearest neighbors in the physical lattice), one can use the circuit shown in Fig.~\ref{fig:experiment}b, to first SWAP a system spin with the intervening ancilla, then apply $U$ to the system-ancilla bond, and then undo the original SWAP. We note that this procedure is limited by the fidelity of the SWAP operations used to implement the virtual layering.

\subheader{Discussion}
So far we have focused on the case of the simplest rational CF phase with $\nu = \log 2$, where we introduced the chiral mutual information, $\chi$MI as a means to characterize and possibly experimentally measure the chiral unitary invariant, $\nu$. This setup works equally well for more general rational CF phases of bosons, with $\nu = \log r$ for arbitrary rational fraction $r$.

In the presence of strong interactions, the periodic drive can also induce dynamical fractionalization, in which the $2d$ bulk develops dynamical Abelian topological order~\cite{po2017radical}. In such fractionalized settings, the chiral unitary invariant can take ``radical" values of the form $\nu = \log \sqrt{r}$ where $r\in \mathbb{Q}_+$ is a rational fraction. In these settings, the Floquet evolution for the system does not simply decompose into commuting edge and bulk pieces~\cite{po2017radical}, and it is presently unclear whether $\chi$MI can be used to directly detect the radical CF invariant. However, in this case, one could instead consider the same setup and evaluate $\chi$MI for two periods, for which the evolution $U(2T)$ becomes rational, and the construction above applies. In this case, one could infer the chiral unitary invariant from: $\nu = \frac 12 \chi[U(2T)]$. \\

\noindent{\it Acknowledgements --} We thank K. Hazzard for insightful conversations. This work was supported by NSF DMR-1653007 (ACP) and was performed in part at Aspen Center for Physics, which is supported by National Science Foundation grant PHY-1607611.

\appendix

\section{Review of formal aspects of Chiral Floquet phases}
\label{app:formal}
\subsection{Effective edge evolution}
In this Appendix, we review the results of~\cite{po2016chiral}, which formulated a precise notion of the effective boundary dynamics for a Floquet MBL system. Specifically, suppose the dynamics are produced by a time-periodic local Hamiltonian, $H(t) = H(t+T) = \sum_r h_a(t)$, where $a$ indexes nearby groups of qubits. The long-time dynamics at time $t = nT +\delta$ is captured by $n = \lfloor t/T\rfloor $ applications of the Floquet unitary, $U_F = \mathcal{T}\left\{e^{-i\int_0^TH(t)dt}\right\}$, followed by micro-motion for time $\delta = t-nT$. Since we have considered a system in which $H(t)$ is MBL in the bulk, then in an infinite 2d plane without boundaries,  we may write the Floquet operator, $U_F = e^{-iH_FT}$, as time-evolution with respect to a static MBL Hamiltonian: $H_F = E(\{\tau^z_i\})$. Here, $\tau^z_i$, are local integrals of motion (LIOM), consisting of qubit operators near site $i$ dressed by a cloud of virtual fluctuations that decay exponentially in distance from $i$, and $E$ is some quasi-local function. 

One can truncate the evolution onto a finite region, $A$, with a boundary, $\d A$, in one of two distinct ways. First, one can truncate the time-dependent Hamiltonian to omit terms residing outside $A$, $H^{(A)}(t) = \sum_{a\subset A}h_a(t)$, which produces a truncated Floquet unitary $U^{(A)} = \mathcal{T}\left\{e^{-i\int_0^TH^{(A)}(t)dt}\right\}$. Second, one can truncate the effective MBL Floquet Hamiltonian by dropping terms in the Taylor expansion of $E$, involving LIOM centered outside of $A$. Denote the resulting truncated MBL Hamiltonian as $H_{F}^{(A)}$. The bulk dynamics produced by $U^{(A)}$ and $e^{-iH_F^{(A)}T}$ at a distance $R$ from $\d A$ are identical to accuracy $e^{-R/\xi}$, where $\xi$ is the localization length. However, only $U^{(A)}$ captures any anomalous topological edge dynamics near $\d A$. 

Comparing these two truncation schemes produces an effective $1d$ edge evolution:
\begin{align}
U_\text{edge} = e^{+iH_F^{(A)}T}U^{(A)}
\end{align}
which is exponentially well localized to $\d A$, and which captures any non-trivial topological edge dynamics. This quantity plays an analogous role to that of the effective boundary field theory for a zero temperature equilibrium topological phase. Such equilibrium boundary field theories are anomalous, in the sense that it cannot emerge as the low energy description of microscopic quantum degrees of freedom living only at the sample boundary. Similarly, for chiral Floquet phases, the effective edge dynamics described by $U_\text{edge}$ can be considered anomalous if it cannot arise from time-evolution under a local $1d$ time-dependent Hamiltonian, $H_{1d}(t)$, acting only on the sample edge.

\subsection{Chiral unitary invariant}
In this section, we review the operator-algebraic definition of the chiral unitary invariant~\cite{gross2012index,po2016chiral}. Consider cutting the $1d$ edge of a $2d$ MBL-Floquet system into two halves, and defining a large region, $L$ to the left, and another, $R$, to the right of one of the cuts. Here, by ``large", we mean that each region is much larger than the Lieb-Robinson velocity at the edge times the driving period. Consider the set of all operators acting on the region $L$. Since any product or linear combination of such operators gives another operator in the set, these sets each form an algebra, $\mathcal{A}_{L/R}$.

To simplify notation, let us temporarily focus on the left. Given an orthonormal basis $\{|i\>\}$ for the states in this region, we can define an orthonormal basis for the operator algebra: $e_{ij} = |i\>\<j|$. This operator basis forms an orthonormal set under the inner product: $\<O,O'\> = \text{tr}O^\dagger O'$ where the trace is taken over all states in the region (e.g. over the basis $\{|i\>\}$). From this inner-product on operators, we can define the overlap of two algebras, $\mathcal{A,B}$, by:
\begin{align}
\<\mathcal{A,B}\> \equiv \frac{\sqrt{D_AD_B}}{D_{A\cup B}}\sqrt{ \sum_{i,j=1}^{D_A}\sum_{k,l=1}^{D_B}\big{|}\<e^{(A)}_{ij},e^{(B)}_{kl}\>\big{|}^2}
\end{align}
where $D_A$ is the dimension of the Hilbert space of region $A$. As defined, this overlap is $1$ for independent (commuting) algebras, and the auto-overlap is $\<\mathcal{A,A}\>=D_A$.

The chiral unitary invariant is then constructed by taking the algebra, $\mathcal{A}_L$, for the region left of the cut, time-evolving it by one period, and measuring its overlap with the algebra, $\mathcal{A}_R$ to the right of the cut: $N_R = \<U\mathcal{A}_LU^\dagger,\mathcal{A}_R\>$. $N_R$ quantifies, how many states worth of information can we deduce about $L$ at $t$ by measuring operators in $R$ at $t=t+T$, or roughly: ``how many states are transferred from $L$ to $R$ during the driving period". Then, one should also compute the number of states transferred from right to left: $N_L = \<U\mathcal{A}_RU^\dagger,\mathcal{A}_L\>$. Neither $N_{R/L}$ are separately quantized. However, their ratio $r = \frac{N_R}{N_L}$ is, and its logarithm defines the chiral unitary invariant:
\begin{align}
\nu = \log \frac{N_R}{N_L} = \log \frac{\<U\mathcal{A}_LU^\dagger,\mathcal{A}_R\>}{\<U\mathcal{A}_RU^\dagger,\mathcal{A}_L\>}
\label{appeq:chiralunitary}
\end{align}

Directly measuring this quantity as formulated would require measuring a large number of operator overlaps -- for a complete set of operators in $\mathcal{A}_{L/R}$, and each operator overlap requires averaging over a complete set of states for that region. The number of such required measurements clearly grows exponentially in the size of the $L/R$ regions, with each involving a multi-spin measurement that is effectively as challenging as the SWAP-operator based entanglement measurement sketched in the main text. Hence, the ancilla degrees of freedom yield a potentially large reduction in measurement complexity, which is exponential in the parameter $v_\text{LR}T$.\\

\subsection{Generic form of edge evolution}
The chiral unitary invariant, $\nu$, of the edge evolution $U_\text{edge}$ is invariant under modifying $U_\text{edge}$ by a finite depth local unitary (FDLU) transformation. As a corollary, we can always write the Floquet evolution operator for the edge of a (rational) CF phase of qubits (or spins-1/2) with $\nu=\log 2$ as pure translation $\tau$ modified by an FDLU transformation:
\begin{align}
U_\text{edge} = U_\text{FDLU}^\dagger \tau U_\text{FDLU} = \(U_\text{FDLU}^\dagger \tau U_\text{FDLU} \tau^\dagger\) \tau
\end{align}
Then, note that $\tau U_\text{FDLU} \tau^\dagger$ is also a $1d$ FDLU (in fact, simply $U_\text{FDLU}$ translated one site to the left). In a bosonic system, any $1d$ FDLU has trivial chiral unitary index, and in the absence of any extraneous symmetries, can be generated by an effective time independent Hamiltonian, which we denote as $H_{1d}$.

We note, in passing, that for rational fermion CF phases, there are intrinsically dynamical topological phases for which we cannot always reduce to time-independent evolution, but we can still write: $U_\text{edge} = \mathcal{T}\{e^{-\int_0^t H_\text{1d}(t')dt'}\}\tau$, and the arguments for the quantization in $\chi$ applies equally well.

\section{Partial entanglement}\label{sec:partialEnt}
In the arguments presented in the main text, we considered CF evolution for states in which the ancillary spins are initially maximally entangled with their system counterparts. Here, we were able to analytically establish that $\chi$ is a quantized invariant that reproduces the chiral unitary invariant, $\nu$. We can also consider starting with an arbitrary amount of entanglement where each system/ancilla pair starts in a state:
\begin{align}
|\psi\> = \cos(\alpha)|\uparrow_s\>|\downarrow_a\> - \sin(\alpha)|\downarrow_s\>|\uparrow_a\>
\end{align}
where $\alpha \in \[0,\frac\pi 4\]$ allows one to continuously adjust the entanglement per ancilla: $s(\alpha) = -\cos^2\alpha\log\cos^2\alpha - \sin^2\alpha\log \sin^2\alpha$, from $0$ ($\alpha =0$), to $\log 2$ ($\alpha = \frac\pi 4$). From numerical simulation, we observe that $\frac{\chi}{s(\alpha)}$ appears to be quantized to $1$ for the chiral Floquet phase (up to finite size corrections), for any value of $\alpha>0$. We note that the (asymptotic) quantization requires taking $\alpha$ to be spatially uniform, for example an uniform gradient of $\nabla\alpha\neq 0$ would relax via a non-topological chiral flow of entanglement, spoiling the quantization of $\chi$.

\begin{figure}[t]
\centering
\includegraphics[width=\columnwidth]{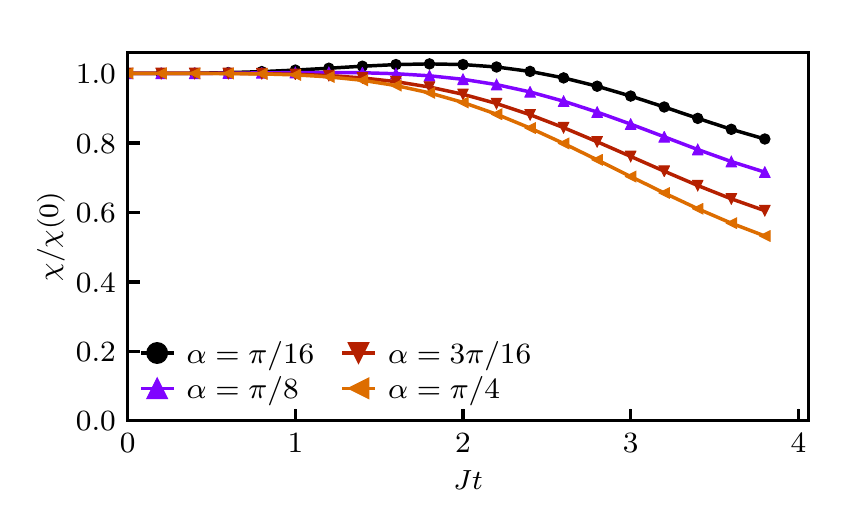}
\vspace{-0.2in}
\caption{
{\bf Partial entanglement --} 
$\chi(t)$ for a state in which each system/ancilla spin pair has variable entanglement characterized by the parameter $\alpha \in [0,\pi/4]$, for $L=12$ and $N_a=6$. Each curve is normalized by the initial value of $\chi(t=0)$. These results give numerical evidence that $\chi(t)$ is quantized for any (uniform) amount of initial system/ancilla entanglement.
}
\label{fig:nonmaxentangle}
\end{figure}

\bibliography{FloqSPTbib}

\end{document}